\documentclass[12pt,a4paper]{article}
\usepackage{latexsym,graphicx}
\newcommand{\eqref}[1]{(\ref{#1})}

\newcommand{\nn}{\nonumber\\}

\newcommand{\x}{{\bf x}}

\renewcommand{\k}{{\bf k}}

\newcommand{\be}{\begin{equation}}
\newcommand{\ee}{\end{equation}}
\newcommand{\bea}{\begin{eqnarray}}
\newcommand{\eea}{\end{eqnarray}}

\newcommand{\subtitle}[1]{\textit{\bf #1.}}
\newcommand{\ok}{$O(2)$ }

\begin{document}
\title{Non-equilibrium Goldstone phenomenon \\
in tachyonic preheating}
\author{
Sz. Bors\'anyi$^{ }$\footnote{mazsx@cleopatra.elte.hu}, 
A. Patk{\'o}s$^{ }$ \footnote{patkos@ludens.elte.hu}
and D. Sexty$^{ }$\footnote{denes@achilles.elte.hu}\\
Department of Atomic Physics\\
E{\"o}tv{\"o}s University, Budapest, Hungary\\
}
\vfill
\maketitle
\begin{abstract}
The dominance of the direct production of elementary Goldstone waves is
demonstrated in tachyonic preheating by numerically determining the evolution
of the dispersion relation, the equation of state and the kinetic power spectra
for the angular degree of freedom of the complex matter field.  The importance
of the domain structure in the order parameter distribution for the
quantitative understanding of the excitation mechanism is emphasized. Evidence
is presented for the very early decoupling of the low-momentum Goldstone modes.
\end{abstract}

\section{Introduction}

The aim of this investigation is to contribute to the systematic exploration of
the transition from the inflationary evolution of the Universe to the standard
cosmological regime.  The numerical analysis was performed in a simple hybrid
inflationary model, in which the inflaton is coupled to a complex scalar field
\cite{linde94,garciab98}. The equations of the fields and of the scale
parameter of Friedmann-Robertson-Walker (FRW) geometry were solved
simultanously. The characteriztics of the transition were studied for a range of
couplings and initial conditions. All choices satisfy the cosmological
constraints entailing a sufficient number of e-foldings during inflation and the
generation of density perturbations compatible with the measured 
cosmic microwave background radiation (CMBR) anisotropy. 

The present investigation is focused on the excitation of Goldstone modes.  In
the literature the decay of global cosmic strings is advocated as the main
source of these particles \cite{davis85,spergel91,yama99,yama00}.  It will be
demonstrated that in the period of tachyonic instability
\cite{felder01,felder01a} a dominant {\it direct} Goldstone generation takes
place.  The importance of Goldstone production was first emphasized by
Boyanovsky {\it et al.} \cite{boyan1,boyan2} in a renormalized large-$N$
approach to the quantum dynamics of the symmetry breaking in the O(N) model.
They have extended their investigation to the FRW geometry in the framework of
the new inflationary scenario \cite{boyan3}.  The present investigation is an
extension of our study of the classical $O(N)$ system in Minkowski metrics
published in Ref. \cite{borsanyi02} to the case of FRW-geometry. In the present
paper also the period of instability is treated classically. Although the time
interval of the instability is rather short it is instructive to study the
transition of the scale parameter of the Universe from the inflationary regime
to a regime dominated by the mixture of weakly interacting species.  We study
the field dynamics under continous variation of the power characterizing
the time dependence of the cosmological scale factor.
  
In hybrid inflationary scenarios the rolling inflaton triggers the condensation
of a complex scalar field thought to be the matter field driving the grand unified theory (GUT) phase
transition. The transition is accompanied by spinodal (tachyonic)
instabilities, which occur for the radial (in the present case,
$O(2)$ invariant) modes.  Rising radial (Higgs) modes excite on their turn
massless angular (Goldstone) modes.  We refer to this phenomenon as the
nonequilibrium Goldstone effect.  It was tested by systematically varying the
lattice spacing that the excitation process is rather insensitive to the
details of the discretization. 

After the tachyonic instability is stopped one observes an excess in the
gradient energy density of the Goldstone degree of freedom relative to the
corresponding kinetic energy density which is argued in \cite{yama99} to be the
signature for the occurrence of finite density of global strings.  Evidence
will be presented for the importance of topological configurations (domain
walls and/or strings) in causing this difference, which is sustained over a
considerable time interval.  Also, a clean radiative equation of state (EoS) is
found for the Goldstone degree of freedom after the virial equilibrium is
reached. After further evolution, however, a clear two-component
separation was observed in the Goldstone power spectra.  Low-(comoving)-$k$
modes decouple from the equilibration processes fairly early and expand further
as a noninteracting massless radiation with a frozen momentum distribution.
High-$k$ modes interact with the massive (Higgs and inflaton) modes.  In this
range we observe quantitative lattice spacing dependence, which does not
challenge the qualitative features of the emerging physical picture. The size
of the decoupled comoving $k$ interval of the Goldstone excitations increases
with time.

In Section 2 the model is presented and the setting of the initial conditions
is discussed. Their choice is crucial to ensure the minimal sensitivity of the
results to the spatial discretization. The most important features of the
excitation process of Goldstone modes, shortly outlined in the previous
paragraph, are discussed in Section 3. A simple estimate will also be given
there for the ratio of the direct Goldstone production relative to the energy
contained in extended objects. The features and the limitations of a
semianalytical model for the excitation of the different modes is discussed in
Section 4.  Section 5 is devoted to the discussion of the late-time expansion.
Conclusions are summarized in Section 6.

\section{Selection of couplings and initial conditions}

The system consists of the real (harmonic) inflaton field $\sigma(\x,t)$ 
and the complex matter field $\phi(\x,t)$, with negative squared mass 
and a quartic stabilizing $O(2)$ invariant self-interaction. Their 
evolution is treated self-consistently assuming a spatially flat FRW 
space-time characterized by its dynamically determined scale function 
$a(t)$.  Metric fluctuations are beyond
the scope of the present study. The equations describing the field 
dynamics of the system are
\bea
\label{field-eq}
0&=&\ddot\sigma(\x,t)+3H\dot\sigma(\x,t)-\Delta
\sigma(\x,t)+m^2_\sigma\sigma(\x,t)+g^2|\phi(\x,t)|^2
\sigma(\x,t),\\
0&=&\ddot\phi(\x,t)+3H\dot\phi(\x,t)-\Delta\phi(\x,t)+m^2\phi+
{\lambda\over
  6}|\phi(\x,t)|^2\phi(\x,t)+g^2\sigma^2(\x,t)
\phi(\x,t).\nonumber
\eea
The FRW equation has the following form:
\bea
H^2-{8\pi\over 3m_{pl}^2}
[\rho_{Higgs}(\x,t)+\rho_{Goldstone}(\x,t)+\rho_{inflaton}(\x,t)]&=&0.
\label{friedmann}
\eea
In Eqs.~(\ref{field-eq}) and (\ref{friedmann}) the Hubble parameter 
$H=\dot a(t)/a(t)$ was introduced and $m_{pl}$ is the Planck mass.  
In the second term of the left hand side of Eq.
(\ref{friedmann}) the expression in the square brackets represents the
microscopical energy density of the physical decomposition of the
fields, to be identified below. The explicit expressions will appear
in Eqs. (\ref{densities1})--(\ref{densities3}).

For the numerical simulation the field equations were rewritten in
conformal time. For instance, in the \ok sector one has
\be
\psi''- \Delta \psi -{ a''\over a} \psi + m^2 a^2 \psi + 
{ \lambda \over 6} |\psi|^2 \psi+ g^2 \Sigma^2 \psi =0,
\ee
with $ \psi=a(t)\phi$ , $ \Sigma=a(t)\sigma$, and both time and space
coordinates are measured in proportion to the scale factor 
[$(d\eta=dt/a(t),  dx=dx_{phys}/a(t))$]. 
Finally, we scale all quantities once more with an appropriate power
of the dimensionless combination $m_da_{init}$, where $m_d$ is an
arbitrary mass unit and $a_{init}$ equals to
the scale factor at the beginning of the simulation
[$a_{init}=a(t_{init})$,  see below]:
\be
\psi_{lat}''- \Delta \psi_{lat} -{ a''\over a} \psi_{lat} + { m^2 a^2
 \over m_d^2 a_{init}^2 } \psi_{lat} + 
{ \lambda \over 6} |\psi_{lat}|^2 \psi_{lat}+ g^2 \Sigma_{lat}^2 
\psi_{lat} =0, 
\ee
The physical quantities are expressed through those simulated on
the lattice as follows:
\be
\phi = \frac{\psi}{a} = \psi_{lat} m_d  { a_{init} \over a } \qquad 
dx_{phys}=dxa={ a \over a_{init}} {dx_{lat} \over m_d}
\qquad dt= a d\eta ={ a \over a_{init}} { d \eta_{lat} \over m_d }
\ee
A convenient choice is to set $m_d=|m|$. In the following, we always use
the powers of $|m|$ as units of measurement.
As one can check, also in the scaled FRW equation only $a/a_{init}$
appears, plus the ratio of the Planck mass and the \ok mass
parameter. Therefore one can study the evolution of 
$a(t)/a_{init}$, independently of the choice of $m_d$.

After this scaling, we discretized the equations of the quantities
labeled by ''{\it lat}'' in a comoving volume $\left(L/a(t)\right)^3$, with
$L=N\delta x_{lat}/|m|\times a(t)/a_{init},~ N=64,128,~ \delta x_{lat}=1$. 
Here $t_{init}$ is the
time instant where we start the numerical solution of Eqs.~(\ref{field-eq}) and
(\ref{friedmann}), which has been conveniently choosen to slightly preceed
the exit point from the inflation. In the plots to be presented below
the time is measured relative to $t_{init}$.  The insensitivity of the
results to the lattice spacing was tested by also employing $\delta
x_{lat}=0.25, 0.5, 0.75$. The conformal time step $\delta \eta$ was chosen
in proportion to the spatial lattice spacing $\delta x$ in the 
range $1/16-1/64$. 

On the other hand, we find for the physical extent of our system
$N\delta{x}_{lat}/m_d\times a(t)/a_{init}\ll H^{-1}(t)$.  
Therefore we actually study only a 
small portion of the volume of the whole Universe. This is different 
than the choice of the lattice spacing in Ref. \cite{yama02}, where the 
system is at least as large as one Hubble volume. Therefore we do not 
expect the string part of the
Goldstone dynamics to be described truly faithfully, but the propagating
quasiparticle excitations are well represented in the simulations. 

We will be able to argue convincingly that in the investigation of the early
appearance of Goldstone modes, the effect of the spatial cutoff will not cause
any finite lattice spacing distortion.  The choice of the lattice constant,
however, is significant for the decay of the Higgs waves into Goldstone
excitations. This process is energetically allowed as long as its comoving
mass $m_Ha(t)/a_{init}$ is smaller than twice the maximal allowed comoving
momentum for the Goldstone waves on the lattice $k_{cutoff}$.  With increasing
redshift one arrives at an artificial stabilization of the Higgs waves.
Therefore, at best, qualitative features of the late time evolution of the
system are expected to be physical.

The matter field started in the symmetric phase, in the close vicinity
of the point
$\phi_0(t_{init})$ $\equiv V^{-1}\int d^3x \phi(\x,t_{init})=0$.
The initial Higgs velocity is $\dot\phi_0(t_{init})=0.$ 
For the homogenous inflaton mode the amplitude and velocity 
values in the  moment $t=t_{init}$
were drawn from the solution of the equations describing
its roll down, started at the Planck scale:
$\sigma_0\equiv V^{-1}\int d^3x \sigma(\x,t=0)=m_{pl}$.

Two important cosmological constraints are to be satisfied. The first 
requires at least $N_{TOT}\sim 60$ e-foldings of the scale factor before
 the critical inflaton field value is reached which terminates 
inflation.  The other constraint stems from the relation of the 
quantum fluctuations of the inflationary period to the density 
fluctuations measured by the COBE experiment
\cite{kolb90}. The numerical method of selecting the couplings without relying 
on the slow-roll approximation was described in some detail in 
Ref.~\cite{sewm02}. 

In the present investigation the GUT scale was chosen 
for the scale of the end of inflation. Therefore for the quantity 
$m_H^2\equiv-2m^2$, variation in the region 
$m_H\sim 10^{(14-15)}$~GeV was allowed. The inflaton-Higgs coupling was 
varied in the interval $g=0.01-0.1$. The value of
the Higgs self-coupling was fixed with the relation $\lambda=3g^2$, 
which is valid if the couplings of the hybrid theory are derived from a 
superpotential \cite{covi01,bellido02}.
The detailed numerical analysis was performed with 
$g=0.1, m_H=8.8\times 10^{14}{\rm
GeV}, m_\sigma= 4.2\times 10^{11}{\rm GeV}, N_{TOT}=60$, and $g=0.01,
m_H=5.5\times 10^{14}{\rm GeV}, m_\sigma=1.4\times 10^{12}{\rm GeV},
N_{TOT}=60$. 

The initial population of the inhomogenous modes imitates the quantum
vacuum. The variation of the lattice spacing changes the size of
the Brillouin-zone, therefore under such variation one would simulate
systems with different cosmological constants. Using
 for the Fourier mode functions the complete orthonormal set 
${1/{(N\delta x_{lat})^{3/2}}}e^{i\k\x}$,
one finds nearly lattice spacing independent energy densities if modes
with $|k|\leq k_{max}$ are filled as follows:
\bea
&
\displaystyle
\sigma_k(t_e)=\sqrt{\frac{1}{{2\omega_\sigma}}}e^{i\alpha_k},
\quad \dot\sigma_k(t_e)=-i\sqrt{\omega_\sigma\over 2}e^{i\beta_k},
\qquad\omega_\sigma^2=k^2+m_\sigma^2,\nn
&
\displaystyle
\phi_{k}(t_e)=\sqrt{\frac{1}{2\omega_\phi}}e^{i\gamma_{k}},
\quad \dot\phi_{k}(t_e)=-i\sqrt{\omega_\phi\over 2}e^{i\delta_{k}},
\qquad\omega_\phi^2=k^2+m^2+g^2\sigma^2(k=0,t_e).
\label{in-cond}
\eea
The initial phases $\alpha_k,\beta_{k},\gamma_k,\delta_k$ were chosen
randomly. It was tested that our conclusions are not sensitive to the
choice of the maximal filled momentum states, e.g $k_{max}\in
(2.,3.)$. In order to have
very accurate equality of the initial energy densities in case of
different lattice spacings, the inhomogenous modes were
filled only up to a lattice spacing dependent
maximal comoving wave number $k_{max}(\delta x_{lat})$. 
For instance, when choosing $k_{max}=2$ for
$\delta x_{lat}=0.5,~N=128$, accurate matching of the initial energy densities
was achieved for $k_{max}=1.84$ 
on the lattice $\delta x_{lat}=1.0,~N=64$. The difference
in $k_{max}$ comes from the distortion of the energy-momentum relation
on lattice. (Be conscious also of the other restriction, that 
during inflation the energy content of the inhomogeneous modes should be
much smaller than the potential energy of the \ok field, e.g. false
vacuum domination).

With this choice we have normalised the energy density to the
same value for any lattice spacing, which leads to finite
final average energy densities when $\delta x_{lat}$ is diminished 
\cite{smit01}. On the other hand, no further renormalization was
necessary to reach lattice spacing independent conclusions concerning
the excitation process.

\section{Direct Goldstone excitation via spinodal instability}

\subtitle{The independent degrees of freedom}
The time evolution of the normalized cross-correlation matrix introduced 
in \cite{borsanyi02} presents evidence that directly after the spinodal
instability is over, one can choose for the three independent degrees of
freedom  the inflaton, the radial O(2) invariant
motion of $r(\x,t)=|\phi(\x,t)|$ and the angular oscillations
$\varphi(\x,t)$, ($\phi=r e^{i\varphi}$). The time evolution of the 
dispersion relation characterizing these degrees
of freedom was calculated from the definition
\be
\omega_k^2={|\dot{X}_k|^2\over |X_k|^2}, \qquad X_k=\sigma_k,r_k, 
(e^{i\varphi})_k,
\label{dispersion}
\ee
and extrapolated to $k=0$ for finding the corresponding masses
\cite{borsanyi02,sewm02}. One obtains for the angular phase 
factor shortly after the spinodal instability mass values which are compatible 
with zero
within the error of the mass determination. This observation justifies the
term ``Goldstone'' for the angular modes. In the following the radial
degree of freedom will be simply referred to as Higgs, and the angular 
component as Goldstone.

\begin{figure}
\begin{center}
\includegraphics[width=12cm]{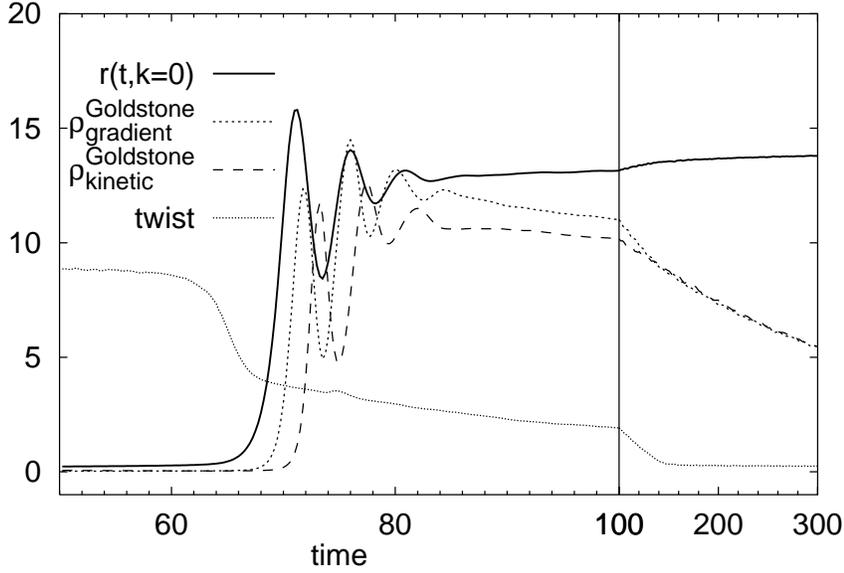}
\end{center}
\caption{Virialization of the Goldstone oscillations, and the
evolution of the average of the integrated twist in $\varphi(\x,t)$, 
calculated
along straight lattice lines parallel to one of the axes.  The 
oscillations of the radial field $\overline{r(\x,t)}^V$ drive the 
variation of the gradient and kinetic energy densities of the Goldstone 
modes  ($g=0.1$, \hbox{$\lambda=3g^2$}, $N=64$).}
\label{kep:bang}
\end{figure}

The phase transition triggered by the inflaton field can be clearly seen 
on the time evolution  
of the homogeneous mode $\overline{r}^V(t)$ of the Higgs field 
(the overline with index ''V'' means spatial averaging).  
The tachyonic 
instability leads to an almost instantanous exponential switch into the 
symmetry broken regime as
shown in Fig.~\ref{kep:bang}. The Higgs field triggers the rise of the
gradient energy of the Goldstone component with a slight delay and
the increase of the Goldstone kinetic energy starts with a further delay. 
The sharp increase in the radial component is terminated by an oscillatory
period, whose frequency is determined by the sum of the classical mass 
square around the minimum $r_0$ of the potential and the space average of 
the Higgs fluctuations: $-2m^2+\lambda\overline{r^2}^V/2$, which is often 
referred to as the Hartree mass. It is clear from the figure that the 
oscillations in the Goldstone kinetic and gradient energies forcefully 
follow the same frequency.

\begin{figure}
\begin{center}
\includegraphics[width=12cm]{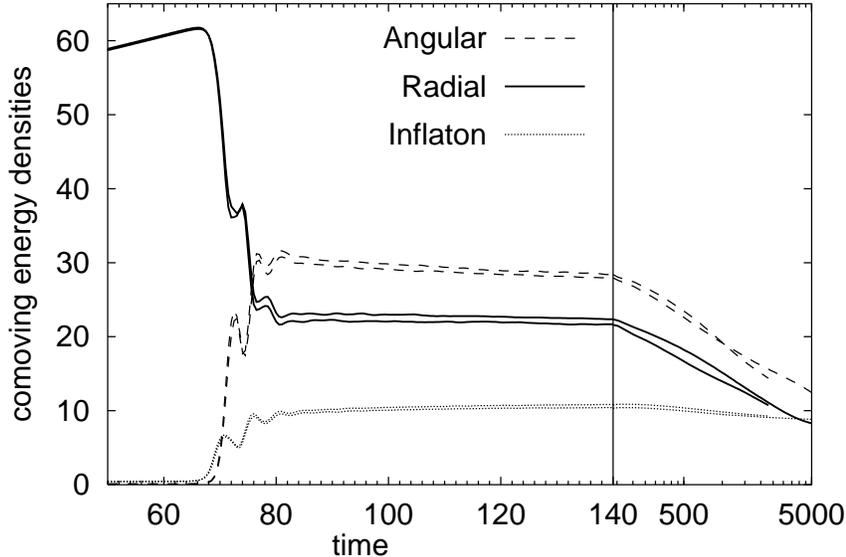}
\end{center}
\caption{Average comoving energy densities
 ($(a^2(t)/a^3_{init})\rho_i(t),~ i=G,H,inflaton$)
 of the independent degrees of freedom
($g=0.1,\lambda=3g^2, N=64, (\delta x_{lat}=1), N=128, (\delta
  x_{lat}=0.5)$). The figure represents the average of 8 runs for $N=64$
 and 4 for $N=128$, each starting with random initial phases. The
  evolution for $t>140$ is plotted on a logarithmic time scale. The
  shorter curves correspond to the solutions on finer spatial lattice.} 
\label{kep:Tstory}
\end{figure}

In Fig.~\ref{kep:Tstory} we analyze the time dependence of the 
average energy densities.  The twin curves, which refer to solutions in 
the same physical volume, but with different lattice spacings illustrate
to what extent the  dynamics of different degrees of freedom
depends on the details of discretization during spinodal instability.
The Higgs component 
starts to vary first, reflecting the instability. The
inflaton and the Goldstone energy densities follow it with 
approximately equal delay. Clearly,  the excitation of  
these modes is driven by the temporal variation of the Higgs field $r(\x,t)$
 (see section 4).  Fig. \ref{kep:Tstory} also suggests that for $g=0.1$ 
the three motions are apparently decoupled from each other 
after the out-of-equilibrium oscillations are damped, each having its 
own nearly constant density.  

The Goldstone field reaches the highest energy density $\sim 30|m|^4$, 
which is slightly higher than the energy density corresponding to the
one-loop estimate of the critical temperature for an 
\hbox{$O(N_{comp}=2)$} model, that is
$\sqrt{72/(N_{comp}+2)\lambda}|m|\sim 25|m|$, 
 with $\lambda=0.03$.  The other two degrees of freedom are much 
colder.
Instant ``freezing'' characterizes the behavior of the inflaton field. It
obtains a rather large squared mass, nearly equal to the Higgs mass, due 
to the
supersymmetric $\lambda -g^2$ relation (e.g.
$m_\sigma^2-6g^2m^2/\lambda\approx -2m^2=m^2_H$). One calculates the 
potential
energy of the inflaton with its ``new'' mass when checking the virial
equilibrium in this degree of freedom. The sudden increase of its full energy
density can be semi-quantitatively
understood to be the result of this mass change (see section 4).

On the right edge of Fig.~\ref{kep:Tstory} the effect of the 
late-time
expansion appears: the energy density of the inflaton field is hardly
varying which corresponds to its nonrelativistic nature. On the
other hand the energy density ratio
 of the Higgs and the Goldstone fields stays nearly constant for $t\le
 5000$. We shall return to the discussion of the apparently 
coupled cooling of the radial and of the angular $O(2)$ components 
in section 5.

\subtitle{The equations of state (EoS)}
We observed that the system stays in a rather stable way deep in the
broken symmetry phase, despite the fact that the Goldstone ``temperature''
is high. One might attempt in such circumstances to characterize
 each species as a noninteracting (decoupled) gas, possessing its own 
equation of state. The local energy densities and pressures have the
following expressions:
\bea
\rho_{Higgs}(\x,t)&=&
\frac{1}{2}\dot{r}(\x,t)^2+\frac{1}{2}(\nabla
r(\x,t))^2+\frac{3}{2}\frac{m^4}{\lambda}+\frac{1}{2}m^2r(\x,t)^2+\frac
{\lambda}{24}r(\x,t)^4,\nonumber\\
p_{Higgs}(\x,t)&=&\frac{1}{2}\dot{r}(\x,t)^2-\frac{1}{6}(\nabla
r(\x,t))^2-\frac{3}{2}\frac{m^4}{\lambda}-\frac{1}{2}m^2r(\x,t)^2-\frac
{\lambda}{24}r(\x,t)^4,
\label{densities1}
\eea
\bea
&
\displaystyle
\rho_{Goldstone}(\x,t)=\frac{1}{2}r(\x,t)^2\dot\varphi(\x,t)^2+\frac{1}{2}
(\nabla\varphi(\x,t))^2,\nonumber\\
&
\displaystyle
 p_{Goldstone}(\x,t)=\frac{1}{2}
r(\x,t)^2\dot\varphi(\x,t)^2-\frac{1}{6}r(\x,t)^2(\nabla\varphi(\x,t))^2,
\label{densities2}
\eea
\bea
&
\displaystyle
 \rho_{inflaton}(\x,t)=\frac{1}{2}\dot\sigma(\x,t)^2+\frac{1}{2}(\nabla\sigma
(\x,t))^2+\frac{1}{2}g^2\sigma(\x.t)^2r(\x,t)^2,\nonumber\\
&
\displaystyle
p_{inflaton}(\x,t)=\frac{1}{2}\dot\sigma(\x,t)^2-\frac{1}{6}
(\nabla\sigma(\x,t))^2-\frac{1}{2}g^2\sigma(\x,t)^2 r(\x,t)^2.
\label{densities3}
\eea

In the equations of state the space averages of the above expressions
appear. Note that the Higgs inflaton interaction term is associated,
with some arbitrariness, exclusively with the inflaton. It is justified
by the realization of the virial equlibrium for this field.

\begin{figure}
\noindent\centerline{\noindent\hbox{\includegraphics[width=9cm]{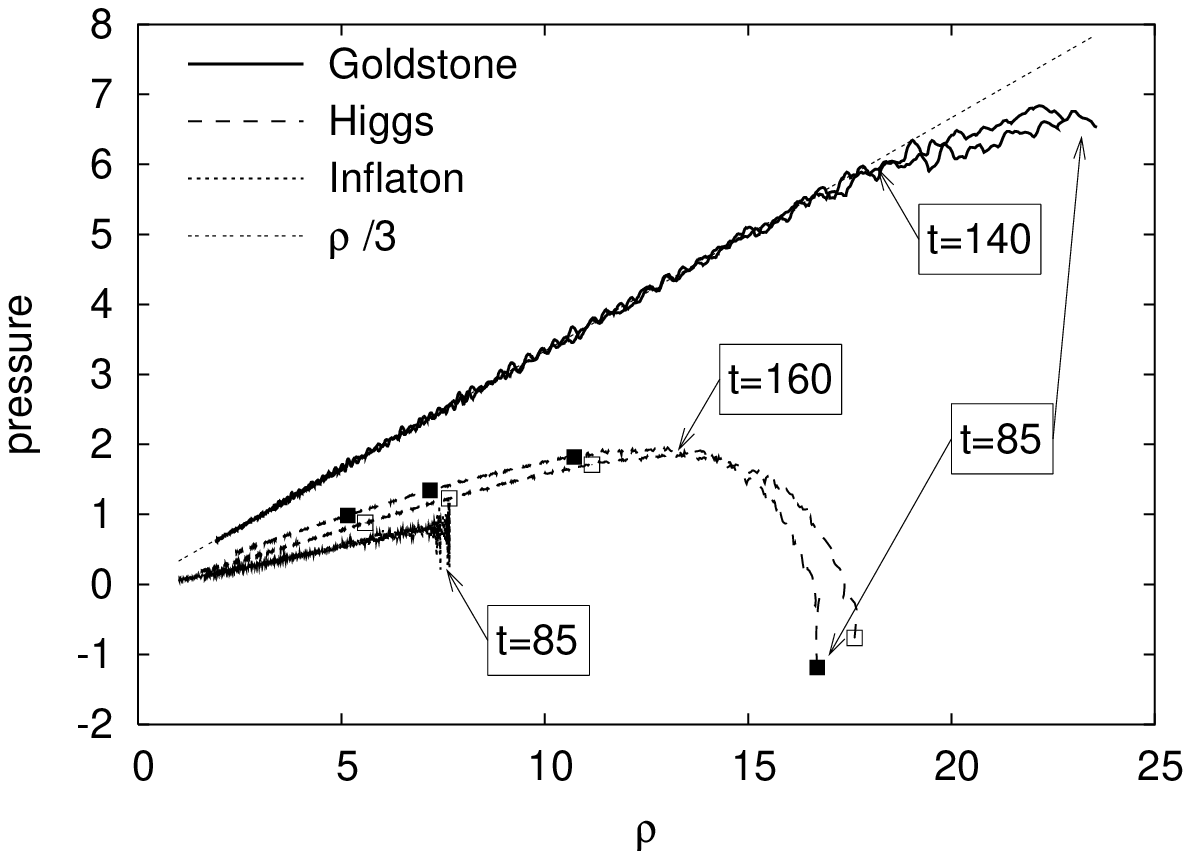}
\includegraphics[width=9cm]{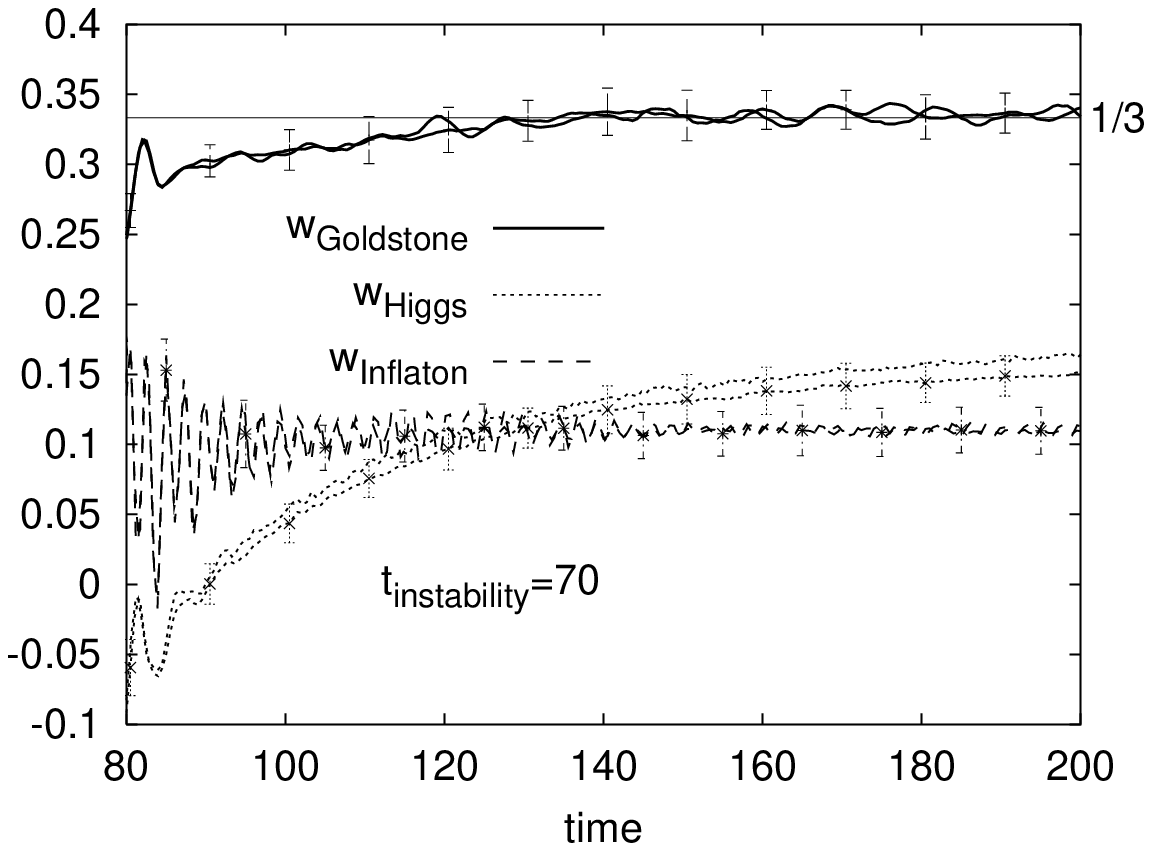}}}
\caption{Equations of state (EoS) for the independent degrees of freedom 
$(g=0.1, \lambda=3g^2)$. Left, the $p(\rho)$ trajectories are shown for the
three independent degrees of freedom. The moments of time,
where the linear regime, characterized by the constant $w=p/\rho$
 sets in are indicated next to the curves. The twin curves display
 results obtained on lattices of equal physical size, but of different
 lattice spacing $(N=64,\delta x_{lat}=1$ (open square), 
$N=128, \delta x_{lat}=0.5$ (full square)$)$. 
The curves represent the average of 20 runs for $N=64$ and 
10 for $N=128$. In the right figure the evolution of the
$w_i=p_i/\rho_i$ ratio for
the different physical degrees of freedom is shown, $(i=G,H, inflaton)$.
}
\label{kep:eos}
\end{figure}

In Fig.~\ref{kep:eos} the EoS of all three degrees of freedom are shown. 
It is clear, that they possess an EoS  from
rather early times, promptly after the large
amplitude $r(t,\x)$ oscillations are damped.They are nearly linear,
of the form $p=w\rho$. The cooling pushes the
system through the points of the EoS at the pace of the expansion, 
therefore
the different points can be labeled by the instant when the system passes
through them.

The inflaton and the Higgs field have nearly the same mass of the order of
 the GUT scale. Therefore we expect them (after the virial ``equilibrium''
 is reached) to follow {\it nearly} the same $p-\rho$ smooth line, 
with the decrease of the energy density due to the expansion.  
Since the inflaton is almost 
decoupled one expects for it $w\approx 0$, and finds at early times $w\sim
 1/10$. The Higgs field starts with a slightly larger slope. We
 note, that the linear regime is reached 
the slowest by the Higgs field, and also its trajectory in the $p-\rho$ plane 
displays noticeable quantitative lattice spacing dependence.

The Goldstone oscillations obey after virialization a perfect radiative 
EoS $(p_G=w_G\rho_G, w_G\approx1/3)$.  For larger energy densities 
($\rho_G\geq 18-23$), that is for earlier times $(85\leq t\leq 140)$, 
one observes a slight deviation from the slope 1/3.  
As one sees in the right hand plot of Fig.\ref{kep:eos} the
curve of $w_G$ determined with help of the
space averages over the Goldstone densities (\ref{densities2}) approaches
$w_{rad}=1/3$ rather smoothly. Practically no
dependence on the lattice spacing can be observed.

This smooth functional form 
 allows a simple quantitative estimate for the composition of the
``Goldstone gas'', if one assumes that it consists of a noninteracting 
mixture of elementary gapless Goldstone waves and of 
nonrelativistic heavy objects composed of coherent configurations
of the angular degree of freedom of the \ok-field. The
measured ratio $\rho/p$ is given then by
$3(1+y),~y=\rho_{\rm{heavy}}/\rho_{\rm{elementary}}$.  This ratio
smoothly approaches zero around $t|m|\sim  140$, till
when the decay of the heavy extended objects will be complete on both the 
$N=64$ and the $N=128$ lattices.
The time dependence of the ratio
$\rho_{\rm{heavy}}/\rho_{\rm full~Goldstone}$ is compatible with an 
exponential
decay with a lattice spacing independent rate: $22(4)~|m|^{-1}$. 
This fit was restricted to the 
range $ 100<t|m|<200$, where the EoS is already well defined and 
the energy
density of the heavy objects is above the noise level. We expect
that these topological objects were created during the tachyonic 
instability,
hence at the time the EoS is stabilized some of them might have 
decayed
already. As an estimate for the initial energy density confined to
topological objects we
extrapolated the exponential time dependence back
 to the moment of the instability ($t\approx 70$) and found that 
(for both lattice spacings realizing the lattice size: $N\delta x_{lat}=64$) 
slightly more than one third of the energy density of the angular motion
was concentrated in heavy
objects and two thirds were carried by massless quasiparticles. A
 polynomial fit to $w_G(t)$ would give an even lower estimate for
 $y(t=70)$. This analysis substantiates
our claim that direct Goldstone production dominates in the 
tachyonic preheating.

One finds complementary information on the Goldstone evolution from  
Fig.~\ref{kep:bang}. The gradient energy density grows higher 
than the kinetic energy as the Goldstone field gets excited. 
This excess reflects
the formation of topologically characterizable extended objects.
In Fig.~\ref{kep:bang} we also showed the average of the integral of
$\delta\varphi(x)$, which is the variation of the angular orientation of 
the complex field from one site to the next one along straight lines 
parallel to the three axes. This integral is very large before the 
instability.  After a sudden drop suffered at the moment of the 
tachyonic instability, it continues
to decay gradually. This evolution goes parallel with the disappearance 
of the gradient energy excess. 

\begin{figure}
\begin{center}
\includegraphics[width=12cm]{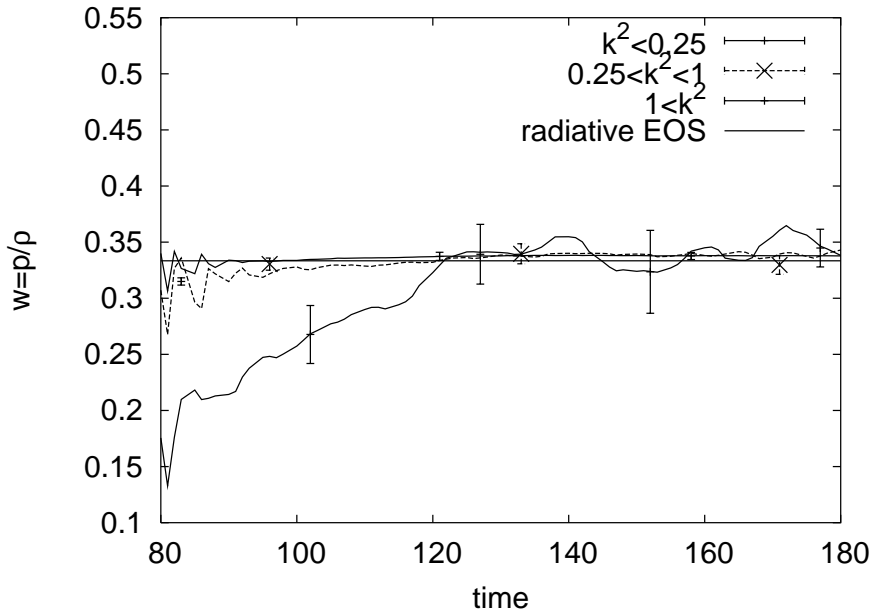}
\end{center}
\caption{The evolution of the EoS of 
the Goldstone oscillators in three different Fourier regions. 
The curves show the average over eight runs for $N=64$, the calculated 
error of the average is signalled by the error bars.} 
\label{wplot}
\end{figure}

In order to deepen the 
understanding of the role played by these topological objects we 
disentangle their contribution to the energy density and the energy 
fraction carried by
the ``elementary'' Goldstone quasiparticles. 
The topological objects contribute only in the low $k$ region of 
the Fourier space. 
 In order to realize this idea, the $k$ space of the Goldstone
degrees of freedom was splitted into three characteriztic regions:
$k^2/|m|^2=[(0,0.25),~(0.25,1.0),(1.0,k_{cutoff}^2)]$. Separate EoS were
fitted in the three regions. In Fig.~\ref{wplot} 
one sees that the deviation from the radiative EoS is localized to the 
lowest $k$ region. (The curve is rather insensitive to the exact choice 
of the values of the separating wave numbers). From a linear fit to
$w_{G,{\rm low k}}(t)$ on the interval $80\leq |m|t\leq 120$, one extrapolates
for $t=70$ $\rho_{\rm heavy}(k^2\leq
0.25|m|^2) /\rho_G(k^2\leq 0.25|m|^2)\approx 0.5$.

\section{On the mechanism of tachyonic mode excitation} 

In hybrid inflationary scenarios the possible mechanisms for the Higgs
field excitation were already discussed at length in the literature
\cite{covi01,bellido02,copeland02}. In this paper we concentrate
on the angular component of the matter field and on the inflaton.

The excitation of the Goldstone and the inflaton field is driven by the 
Higgs field. It turns out that the gradient energy density of $r(\x,t)$ 
is about five times smaller than its kinetic energy density during the 
instability interval, therefore it is reasonable to replace $r^2(\x,t)$ 
in the field equations of the Goldstone and of the inflaton fields by 
its spatial average, $\overline{r^2}^V$.

The analysis is particularly simple for the inflaton, because in this
approximation the equations of its spatial Fourier components are linear.
According to the proposed model the symmetry breaking simply increases 
the effective squared masses of these uncoupled oscillators
\be
\ddot{\sigma}_{\k}+\omega^2_{\k}(t)\sigma_{\k} =0, \qquad \omega^2_{\k}
(t)=\k^2+m_\sigma^2+g^2\overline{r^2}^V(t).
\ee
Since we are interested in the excitation mechanism, which is rather
unaffected by the overall expansion ($H^{-1}\gg \Delta 
t_{\rm reheating}$) we dropped the derivatives of the scale factor in 
these equations.

One can construct the energy balance of each mode by integrating the
equation
\be
\frac{dE_\k}{dt}=\frac{1}{2}\frac{d\omega_\k^2}{dt}\sigma^2_\k,
\ee
which yields
\be
E_{\k}(t)-E_{\k}(0)={g^2\over 2}\int_0^tdt'
{d\overline{r^2}^V(t')\over dt'}
\sigma^2_{\k}(t').
\ee
If one sums up the equations for all $\k$, one finds
\be
E_{{\rm inflaton}}(t)-E_{{\rm inflaton}}(0)={g^2\over 2}\int_0^tdt'
{d\overline{r^2}^V(t')\over dt'}\overline{\sigma^2}^V(t').
\ee

This result may be easily compared to the full numerical solution.
The agreement is quite spectacular when one 
fills the Brillouin-zone completely, and deteriorates when $k_{max}$ is 
decreased. The approximation based on homogenous Higgs fluctuations
 is less appropriate for the lower-$k$ inflaton modes.

A similar approximate construction might be attempted
 for the Fourier transform of the angular variable, which leads
to the equation:
\be
\ddot\varphi_{\k}(t)+2{d\over
  dt}\overline{\ln r(\x,t)}^V\dot\varphi_{\k}+k^2\varphi_{\k}=0.
\label{angosc}
\ee
This is a set of equations for
independent oscillators damped by a common friction.
An equation  Eq.~(\ref{angosc}) could in principle account for the 
delay in
the excitation of the Goldstone kinetic energy (see Fig.~\ref{kep:bang}) 
since
the short but strong common friction effect nearly stops the initially
independent angular oscillation modes $(\dot\varphi_{\k}\approx 0)$ 
in the same instant. When $\overline {\ln r}^V$ is stabilized, the
friction disappears and all oscillators 
start to move with equal phase angle.

 Both the friction
coefficient ${d\over dt}\overline{\ln r(\x,t)}^V$ and the initial 
conditions
for $\varphi_{\k}$ were taken from the numerical solution of the full 
dynamics. The energy density resulting from the solution of Eq.~(\ref{angosc}) 
produced much less excitation in the Goldstone modes than one observes 
in the full solution of Eq.~(\ref{field-eq}). 
This forces us to conclude that the inhomogeneity of
$r(\x,t)$ plays important role also in the Goldstone excitation. 
This conclusion seems to depend rather
sensitively on the value of the couplings $\lambda, g^2$. The smaller is
$\lambda$ the more important is the inhomogenous contribution to the Higgs
kinetic spectra already in the first oscillation period after the spinodal
instability.

\section{Expansion and late-time cooling}

The variation of the cosmological scale factor directly after the
instability reflects radiation domination.  We find by a high
quality power law fit to its numerical evolution the behavior
\begin{equation}
a(t)\sim (t-{\rm const.})^\gamma,\qquad \gamma = 0.54(2),
\label{gammavalue}
\end{equation}
where the error is estimated from the standard deviation of the fitted 
$\gamma$ values when the time interval $t|m|=100-1000$ is splitted into 
several shorter intervals.
One may note that the value of $\gamma$ in the time interval (100-2000)
is reasonably independent of the lattice constant. The variation of the scale
factor is $a(|m|t=2000)/a_{\rm init}\approx 2.5$. The lattice artefacts
appear for $|m|t>2000$.
This exponent, however, shows a time 
dependence on longer time scales as it approaches 
$\gamma_{\rm matter}=2/3$.
This experience shows that special care must be taken if one wishes to 
enforce a simple power law behavior for $a(t)$
over an extended time interval as it 
was done in Refs.~\cite{yama99,yama00,copeland02}. 

If the equations of state are linear, then the actual rate of expansion
conforms to the equation of state of the mixture system: 

\begin{eqnarray}
&
\displaystyle
p_{\rm full}(t)=w_{\rm full}(t)\rho_{\rm full}(t), \qquad
\gamma(t)=\frac{2}{3(1+w_{\rm full}(t))}, \\
&
\displaystyle
p_{\rm full}=p_{\rm Goldstone}+p_{\rm Higgs}+p_{inflaton},\qquad
\rho_{\rm full}=\rho_{Goldstone}+\rho_{Higgs}+\rho_{inflaton}.\nonumber
\label{gammaformula}
\end{eqnarray}

This $w_{\rm full}$ coefficient is a subject of continuous shift from 
$w_{\rm full}\approx 0.23(1)$ towards zero, which would mean the ideal 
matter dominated
EoS. For the time interval of the fit Eq.~(\ref{gammavalue}) the 
time-averaged value $\overline{w_{full}}=0.22(2)$ agrees well with the 
fitted $\gamma$ exponent.

For completely decoupled field components one would expect
\be
3(1+w_i(t))=-\frac{d\ln\rho_i(t)}{d\ln a(t)},\quad i={\rm Goldstone,
Higgs, inflaton}
\label{diffw}
\ee
with an (almost) time independent $w_i$ value.  They should agree with 
the corresponding coefficients of the different EoS, if the temporal
evolution consists simply of the expansion of the noninteracting gas
components. Indeed, this is the case of the inflaton, but the $O(2)$
sector shows rather large deviations.  From Eq.(\ref{diffw}) one finds
that the Goldstone energy density on the average decreases more
slowly with $a(t)$ than expected for a massless radiation.
A similar comparison reveals that the rate of cooling of the Higgs mode is
faster than for a nonrelativistic gas.
The powers $\gamma_i(t)$ vary with time quite strongly.
These tendencies indicate an energy transfer to the angular motion in
the $O(2)$ configurational space from the radial one.
The situation is puzzling, since
the average energy density of the Goldstone modes is higher than that of
the radial oscillations.

\begin{figure}[ht]
\begin{center}
\includegraphics[width=12cm]{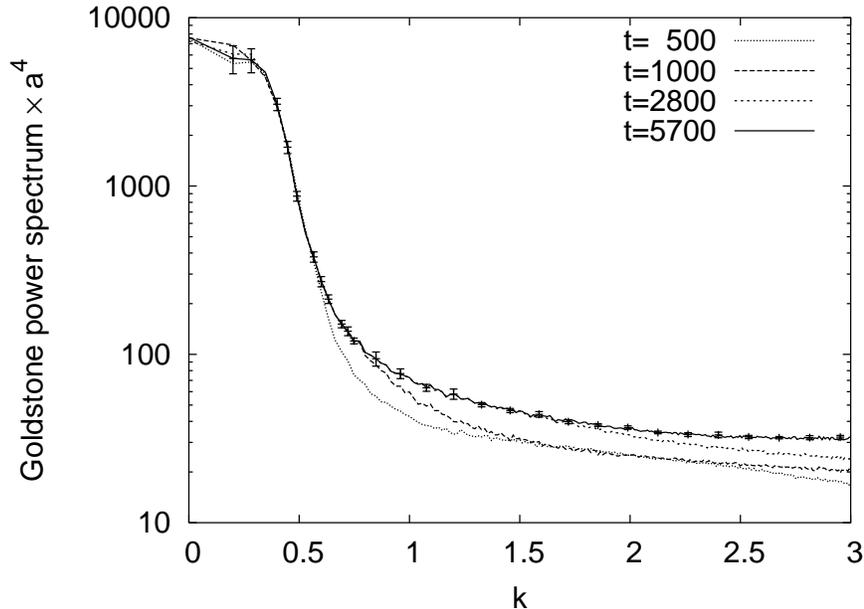}
\end{center}
\caption{Time evolution of the kinetic
energy spectra of the Goldstone excitations
displayed as a function of the comoving wave number.
Average over 8 runs with typical standard deviations as error 
bars for one of the spectra are shown. 
The rescaled late time spectra agree within
the standard deviation in an interval $(0,k_{\rm lim})$
which increases with time ($g=0.1, N=64$).}
\label{kep:spectra}
\end{figure}

The investigation of the kinetic power spectra provides much more detailed
information. One finds that the Higgs and inflaton degrees of freedom
quickly approach classical equipartition as suggested by their EoS.
The power spectra of the Goldstone degree of freedom reveal interesting
regularities.  Fig.~\ref{kep:spectra} shows the Goldstone
energy spectra multiplied by $a^4(t)$ 
at late times (well within the regime of virial equilibrium). The 
shape near the origin is very similar to what was found in Minkowski metric
\cite{borsanyi02}.  The durable deviation from equipartition appears as a
result of anomalously slow relaxation in the lowest comoving
$k$ region, frozen in an almost instant decoupling after the
tachyonic instability. This is demonstrated by the fact that this part
of the spectra shows perfect $a^{-4}(t)$ scaling.  
The height and the width of this decoupled out-of equilibrium spectra in
the low-$k$ region is invariant under a factor-of-two change of the
lattice spacing, therefore the early decoupling is certainly a
physical effect.
 
The Goldstone energy density in the high-$k$ region is much lower but it 
gradually increases and approaches some sort of equipartition.
This rise is apparently fed by the uniform 
decrease of the power of the radial (Higgs) motion. The $k_{\rm lim}$ value 
separating the region where the spectral power of the Goldstone is
already stationary from the part which is still increasing
 is shifted gradually towards the cutoff.
The final uniform level of excitation in the high-$k$ region is lower
for finer lattices of the same physical size. This is easy to interpret
by noticing that the same initial energy density is now ditributed when
equipartition is approached among different numbers of degrees of
freedom.

The synchronized variation in the Higgs and Goldstone power spectra
 described above has a simple physical explanation. The
massive Higgs waves can decay into energetic pairs of elementary
Goldstone modes irreversibly, even if their temperature is lower.
 In the language of conformally transformed variables the
radial mass term scales with $a(t)$.
We find $k_{\rm lim}(t)a_{init}/a(t)$ 
to be approximately constant, and its value
corresponds to the equality:
$2k_{\rm lim}(t)a_{init}/a(t)\approx m_{H}$. The main distinction
between the two regions of Goldstone modes consists in the circumstance,
that the low-(comoving)-$\k$ modes do not receive energy input from
Higgs-to-Goldstone pair
creation. As $k_{\rm lim}$ sweeps through the Brillouin-zone completely
decoupled Goldstone modes are left behind. The physics of the system
becomes seriously distorted at times when $m_{H}a(t)/a_{init}$ 
approaches $2k_{\rm cutoff}$, beyond which the Higgs waves will 
be artificially stabilized. This lattice spacing sensitivity is
manifested in the level of excitation only for $|\k|>1$.

\section{Conclusions}

In this study we focused on the very early stage of the
field evolution following the tachyonic instability occuring in simple
realizations of the hybrid inflationary scenario. The results were
shown overwhelmingly insensitive to the choice of the lattice spacing
if the inhomogenous
modes were filled initially only up to a maximal wave number. Its value 
was chosen so, that the energy density of the system
was kept the same for different lattice spacings. 
Lattice spacing independence of the main observables was
achieved without imposing any further normalization condition.

We found that the
direct production of Goldstone excitations is very efficient. 
By a thorough analysis of the low-$k$ part of the Goldstone spectra and 
its contribution to the equation of state one can separate the elementary 
waves from the extended objects (strings) formed from coherent Goldstone
configurations. The decay of the strings can be followed through the 
temporal variation of the corresponding 
$w_G=p_{G}/\rho_{G}$
ratio. The smaller is the selfcoupling $\lambda$ the longer the 
string-network lives.

The simple-minded model of the motion of Goldstone  and inflaton modes,
consisting of independent oscillators moving in the background of homogenous
time-dependent Higgs fluctuations provides a semiquantitative 
interpretation of the excitation of these degrees of freeedom.

Curiously, we also found an interesting mechanism for energetic reheating in
our investigation which still avoids the restoration of the symmetry.  It might
happen that some kind of strongly out of equilibrium dynamics (in our case the
tachyonic instability) excites a certain degree of freedom much more
efficiently (in our case the Goldstone's) than the order parameter.  If their
interaction is weak relative to the Hubble expansion rate, decoupling occurs
very early and the matter field cannot climb out from the symmetry breaking
minimum. The Fourier power spectra of the Goldstone mode is frozen in a
strongly out of equilibrium shape, displaying the enhancement of low-$k$ modes.
One has to question if the decoupling is sensitive to the choice of the lattice
spacing. We have checked that the efficient Goldstone excitation is followed by
quick decoupling preserving the same shape for the power spectra in the
$|\k|<1$ region for $\delta x_{lat}=0.25, 0.5, 0.75, 1.0$.

During the later transition period from radiation towards matter
domination the decay of the Higgs waves into Goldstone modes can be
observed.  The spectra 
did show evidence for a lattice spacing dependence in the
$|\k|>1$ region.

The features of the late time dynamics in the O(2) sector, analyzed above can
be summarized in a spectral variant of the system of coupled equations, which
describes the variation of the energy densities contained in (radiation like)
light modes coupled to a heavy degree of freedom \cite{kolb90}.  The variation
in conformal time of the comoving Goldstone mode energies is affected by the
decay of the Higgs waves into Goldstone particles. By the assumption that the
momentum distribution of the Higgs degree of freedom obeys at each instant the
equipartition rule, one conjectures the following kinetic equations:

\bea
{d\over d\eta}(\rho_G(\k)a^4(t))&=&\Theta \left(|\k|-\frac{1}{2}m_Ha(t)\right)
\tau^{-1}\frac{3}{4\pi}\frac{1}{k_{cutoff}^3}\rho_Ha^4(t),\nonumber\\
{d\over d\eta}(\rho_Ha^3(t))&=&-\left(1-\frac{1}{8}\frac{(m_Ha(t))^3}
{k_{\rm cutoff}^3}\right)
\tau^{-1}\rho_Ha^4(t)
\eea
($\tau^{-1}$ is the decay rate of the Higgs waves).

The next stage of our project is to extend the investigation to the case 
of gauged models of hybrid inflation
\cite{copeland02,skullerud02,garcia03}, 
which differs in very important aspects from the models where
Goldstone modes appear.

\section*{Acknowledgements}

The authors enjoyed very informative discussions with A. Jakov\'ac and 
Zs. Sz{\'e}p. Remarks by M. Hindmarsh at SEWM'02 are gratefully 
acknowledged.  
This research was supported by the Hungarian Research Fund (No. T037689).


\begin{thebibliography}{9}

\bibitem{linde94}A.D. Linde, Phys. Rev. {\bf D49} (1994) 748  
\bibitem{garciab98} J. Garcia-Bellido and A.D. Linde, Phys. Rev. {\bf
  D57} (1998) 60557
\bibitem{davis85} R.L. Davis, Phys. Rev. {\bf D32} (1985) 3172
\bibitem{spergel91} D.N. Spergel, N. Turok, W.H. Press and B.S. Ryden,
  Phys. Rev. {\bf D43} (1991) 1038
\bibitem{yama99} M. Yamaguchi, Phys. Rev. {\bf D60} (1999) 103511
\bibitem{yama00} M. Yamaguchi, J. Yokoyama and M. Kawasaki,
Phys. Rev. {\bf D61} (2000) 061301(R)
\bibitem{felder01}  G. Felder, J. Garcia-Bellido, P.B. Greene,
  L. Kofman, A.D. Linde and I. Tkachev, Phys. Rev. Lett. {\bf 87}
  (2001) 011601
\bibitem{felder01a} G. Felder, L. Kofman and A.D. Linde,
  Phys. Rev. {\bf D64} (2001) 123517
\bibitem{boyan1} D. Boyanovsky, H.J. de Vega, R. Holman, D.S. Lee, and
  A. Singh, Phys. Rev {\bf D51} (1995) 4419
\bibitem{boyan2} D. Boyanovsky, R. Holman and J.F.J. Salgado,
  Phys. Rev. {\bf D54} (1996) 7570
\bibitem{boyan3} D. Boyanovsky, D. Cormier, H.J. de Vega and S. Prem
  Kumar, Phys. Rev. {\bf D57} (1998) 2166
\bibitem{borsanyi02} Sz. Bors\'anyi, A. Patk{\'o}s and D. Sexty,
  Phys. Rev. {\bf D66} (2002) 025014
\bibitem{yama02} M. Yamaguchi and J. Yokoyama, hep-ph/0210343
\bibitem{kolb90} E.W. Kolb and M. Turner, {\it The Early Universe}, 
Addison-Wesley 1990, New York
\bibitem{covi01} T. Asaka, W. Buchm\"uller and L. Covi, Phys. Lett. 
{\bf B510} (2001) 271
\bibitem{bellido02} J. Garcia-Bellido, M. Garcia Perez and 
A. Gonzalez-Arroyo, hep-ph/0208228
\bibitem{sewm02} Sz. Bors\'anyi, A. Patk{\'o}s and D. Sexty,
hep-ph/0301117, to appear in Procs. of SEWM'02, Sept. 29-Oct. 2 
2002, Heidelberg, ed. M.G. Schmidt
\bibitem{smit01} M. Sall{\'e}, J. Smit and J.C. Vink, hep-ph/0112057
\bibitem{copeland02} E.J. Copeland, S. Pascoli and A. Rajantie,
Phys. Rev. {\bf D65} (2002) 103517
\bibitem{skullerud02} J. Skullerud, J. Smit and A. Tranberg,
  hep-ph/0210349, to appear in Procs of SEWM'02, Sept 29-Oct. 2 2002,
  Heidelberg, ed. M.G. Schmidt
\bibitem{garcia03} J. Garcia-Bellido, M. Garcia Perez and
  A. Gonzalez-Arroyo, hep-ph/0304285
\end{thebibliography}
\end{document}